\documentclass[a4paper,fleqn]{cas-dc}

\usepackage[numbers]{natbib}
\usepackage{epsfig}
\usepackage{bm}
\usepackage{amssymb}
\usepackage{amsmath}
\usepackage{color}
\usepackage{subfigure}
\usepackage{multirow}
\usepackage{threeparttable}

\newcommand{\sNN}{$\sqrt{s_{\rm NN}}$ }
\newcommand{\GeVc}{GeV/$c$}

\newcommand{\pT}{$p_{\rm T}$}
\newcommand{\dz}{$\rm D^{0}$}
\newcommand{\dpm}{$\rm D^{\pm}$}
\newcommand{\ds}{$\rm D_{s}$}
\newcommand{\lc}{$\rm\Lambda_{c}$}
\newcommand{\RAA}{R_{\rm AA}}
\newcommand{\fAA}{f_{\rm AA}}
\newcommand{\fpp}{f_{\rm pp}}
\newcommand{\cte}{\rm c\rightarrow e}
\newcommand{\bte}{\rm b\rightarrow e}
\graphicspath{{figs/}}

\def\tsc#1{\csdef{#1}{\textsc{\lowercase{#1}}\xspace}}
\tsc{WGM}
\tsc{QE}
\tsc{EP}
\tsc{PMS}
\tsc{BEC}
\tsc{DE}

\begin{document}
\let\WriteBookmarks\relax
\def\floatpagepagefraction{1}
\def\textpagefraction{.001}
\shorttitle{Charm and beauty isolation from heavy flavor decay electrons in Au+Au collisions at $\sqrt{s_{\rm NN}}$ = 200 GeV at RHIC}
\shortauthors{Fan Si et~al.}

\title [mode = title]{Charm and beauty isolation from heavy flavor decay electrons in Au+Au collisions at $\sqrt{s_{\rm NN}}$ = 200 GeV at RHIC}

\author[1]{Fan Si}
\author[1]{Xiao-Long Chen}
\author[1]{Long Zhou}
\author[1]{Yi-Fei Zhang}\cormark[1]\ead{ephy@ustc.edu.cn}
\author[1]{Sheng-Hui Zhang}
\author[1]{Xin-Yue Ju}
\author[1]{Xiu-Jun Li}
\author[2]{Xin Dong}
\author[2,3]{Nu Xu}
\address[1]{State Key Laboratory of Particle Detection and Electronics, University of Science and Technology of China, Hefei 230026, China}
\address[2]{Lawrence Berkeley National Laboratory, Berkeley, CA 94720, USA}
\address[3]{Institute of Modern Physics of CAS, Lanzhou 730000, China}

\cortext[cor1]{Corresponding author}

\begin{abstract}
We present a study of charm and beauty isolation based on a data-driven method with recent measurements on heavy flavor hadrons and their decay electrons in Au+Au collisions at $\sqrt{s_{\rm NN}}$ = 200 GeV at RHIC. The individual electron $p_{\rm T}$ spectra, $R_{\rm AA}$ and $v_2$ distributions from charmed and beauty hadron decays are obtained. We find that the electron $R_{\rm AA}$ from beauty hadron decays ($R_{\rm AA}^{\rm b\rightarrow e}$) is suppressed in minimum bias Au+Au collisions but less suppressed compared with that from charmed hadron decays at $p_{\rm T}$ $>$ 3.5 GeV/$c$, which indicates that beauty quark interacts with the hot-dense medium with depositing its energy and is consistent with the mass-dependent energy loss scenario. For the first time, the non-zero electron $v_2$ from beauty hadron decays ($v_2^{\rm b\rightarrow e}$) at $p_{\rm T}$ $>$ 3.0 GeV/$c$ is observed and shows smaller elliptic flow compared with that from charmed hadron decays at $p_{\rm T}$ $<$ 4.0 GeV/$c$. At 2.5 GeV/$c$ $<$ $p_{\rm T}$ $<$ 4.5 GeV/$c$, $v_2^{\rm b\rightarrow e}$ is smaller than a number-of-constituent-quark (NCQ) scaling hypothesis. This suggests that beauty quark is unlikely thermalized and too heavy to be moved in a partonic collectivity in heavy-ion collisions at the RHIC energy. 
\end{abstract}


\maketitle

\section{Introduction}

The pursuit of Quark-Gluon Plasma (QGP) is one of the most interesting topics in strong interaction physics~\cite{QGP1, QGP2, QGP3}. Recent experimental results from Relativistic Heavy-Ion Collider (RHIC) and Large Hadron Collider (LHC) support that a strongly coupled QGP matter (sQGP) has been created in ultra-relativistic heavy-ion collisions~\cite{QGPexp1, QGPexp2, QGPexp3, QGPexp4}. Studying the properties of the QGP matter and understanding its evolution in the early stage of the collisions are particularly helpful for broadening our knowledge of the early born of the universe. 

Heavy quark (charm and beauty) masses, different from those of light quarks, are mostly coming from initial Higgs field coupling, which is hardly affected by the strong interactions~\cite{BM}. Thus heavy quarks are believed to be produced predominantly via hard scatterings in the early stage of the collisions and sensitive to the initial gluon density. And their total production yields can be calculated by perturbative-QCD (pQCD)~\cite{charm} and are number of binary nucleon-nucleon collisions ($N_{\rm coll}$) scaled. Theoretical calculations predict that the heavy quark energy loss is less than that of light quarks due to suppression of the gluon radiation at small angles due to the quark mass. The beauty quark mass is a factor of three larger than the charm quark mass, thus one would expect less beauty quark energy loss than charm quark when they traverse the hot-dense medium created in the heavy-ion collisions~\cite{Qeloss1, Qeloss2, Qeloss3}. Experimentally, the nuclear modification factor ($\RAA$), which is defined as the ratio of the production yield in A+A collisions divided by the yield in $p$+$p$ collisions scaled by $\langle N_{\rm coll}\rangle$, is used to extract the information of the medium effect, such as the parton energy loss~\cite{RAAXNW}. Recent measurements on the $\RAA$ of open charm hadrons and leptons from heavy flavor (HF) hadron decays show strong suppression at high transverse momenta (\pT), and with a similar magnitude as light flavor hadrons, which indicates strong interactions between charm quark and the medium~\cite{STARD1, STARD3, ExpE1, ExpE2}. However, due to technique challenges, most of the electron measurements are the sum of the products from HF hadron decays without charm and beauty contributions isolated. Recently, with the help of vertex detectors, some of the experiments at RHIC have extracted the charm and beauty contributions from the heavy flavor electron (HFE) measurements but with large uncertainties~\cite{PhecbE, HFTTemplate}. The beauty quark production from semileptonic decay channels has been measured at higher energies at LHC~\cite{LHCb2e}. However, due to the different temperature and system density, the behavior of beauty quark could be different at the RHIC energy. It is also worthy of observing the collision energy dependence of the beauty quark production.

Naively, heavy quarks are too heavy to be pushed moving together with the collective flow during the expansion of the partonic matter unless the interactions between heavy quarks and surrounding dense light quarks are strong and frequent enough. After sufficient energy exchange, the system could reach thermal equilibrium. Therefore, heavy quark collectivity could be an evidence of heavy quark thermalization. The heavy quark elliptic flow, defined as a second harmonic Fourier coefficient ($v_2$) of the azimuthal distribution of particle momenta~\cite{FlowArt}, is proposed to be an ideal probe to the properties of the partonic matter, such as the thermalization, intrinsic transport parameters, drag constant and entropy~\cite{charm, Moore, Andronic, SUBATECH, TAMU1, TAMU2, LBT, PHSD}. Apparently, measuring the charm and beauty quark $v_2$ separately is crucial to constrain the diffusion parameters extracted from quenched lattice QCD~\cite{lQCD1, lQCD2}. In particular, the beauty quark mass is about three times larger than the charm quark mass and the final state behaviors of the two quarks could be different. Unfortunately we are very ignorant of that. Up to date, there are many measurements of HF hadron spectra and $v_2$, but most of them are for charmed hadrons or electrons from HF hadron decays. Some attempts on the separation of charm and beauty contributions in HF decay electrons are only for their momentum distributions. There is no measurement on the beauty quark $v_2$ either in hadronic decays or indirect electron channels at RHIC.

\section{Analysis Technique and Results}

\subsection{Spectra and $\RAA$}

We have developed a data-driven method to isolate charm and beauty contributions from the inclusive HFE spectrum based on the most recent open charm hadron measurements in minimum bias (Min Bias) Au+Au collisions at \sNN = 200 GeV at RHIC. Taking the advantage of the Heavy Flavor Tracker (HFT), the STAR experiment has achieved precision measurements at mid-rapidity ($\left|y\right|$ $<$ 1) on \pT\ spectra of inclusive (prompt and non-prompt) \dz-mesons~\cite{STARD1, STARD3} at 0 $<$ \pT\ $<$ 10 \GeVc, as well as other charmed hadrons (\dpm~\cite{zhou2017measurements}, \ds~\cite{zhou2017measurements} and \lc~\cite{Lcspectra}) at 2 \GeVc\ $\lesssim$ \pT\ $<$ 8 \GeVc. Non-prompt \dz\ (from beauty hadron decays) contributes about 5\% at \pT\ $<$ 8 \GeVc~\cite{b2D} and at higher \pT\ where there is no measurement, the fixed-order next-to-leading log (FONLL) prediction was applied for extrapolation and to be about 10\%~\cite{charm}. The parameterized \dz\ spectrum is extrapolated up to \pT\ = 20 \GeVc\ due to the negligible electron yield from \dz\ decays at $p_{\rm T}^{\rm D^0}$ $>$ 20 \GeVc. The parameterized uncertainties include three parts: a) 1-$\sigma$ band of the \dz\ spectrum by fitting with a Levy~\cite{wilk2000interpretation} function with uncorrelated statistical uncertainties; b) Half of the difference between Levy and power-law~\cite{adamczyk2012measurements} fits; c) For correlated systematic uncertainties the spectrum is scaled to upper and lower limits. The total uncertainty is then quadratically summed from above three components. The \ds\ spectrum at 10$-$40\% centrality is parameterized in the same way, since there is no clear centrality dependence observed based on the current precision. The uncertainty of \dz\ (\ds) \pT\ spectrum within 0 $-$ 10 \GeVc\ is 10.7\% (15.7\%) at low \pT\ up to 44.2\% (49.1\%) at \pT\ = 10 \GeVc. The \dpm\ spectrum is obtained by scaling the \dz\ spectrum with a constant (0.429 $\pm$ 0.038), which is fitted from the yield ratio of \dpm~\cite{zhou2017measurements} divided by \dz~\cite{STARD3}, since there is no clear \pT\ dependence observed. The \lc\ spectrum at 10$-$80\% centrality is fitted and extrapolated down to zero \pT\ and up to \pT\ = 10 \GeVc\ with the measured \dz\ spectrum multiplying different model calculations on the yield ratio of $\rm\Lambda_c/D^0$~\cite{ghosh2014diffusion, zhao2018sequential, ko, rapp}. The uncertainty of \lc\ is 29.7\%$-$77.7\%, which is mainly from the average of the four models.

\begin{figure}
\centering
\includegraphics[width=\linewidth]{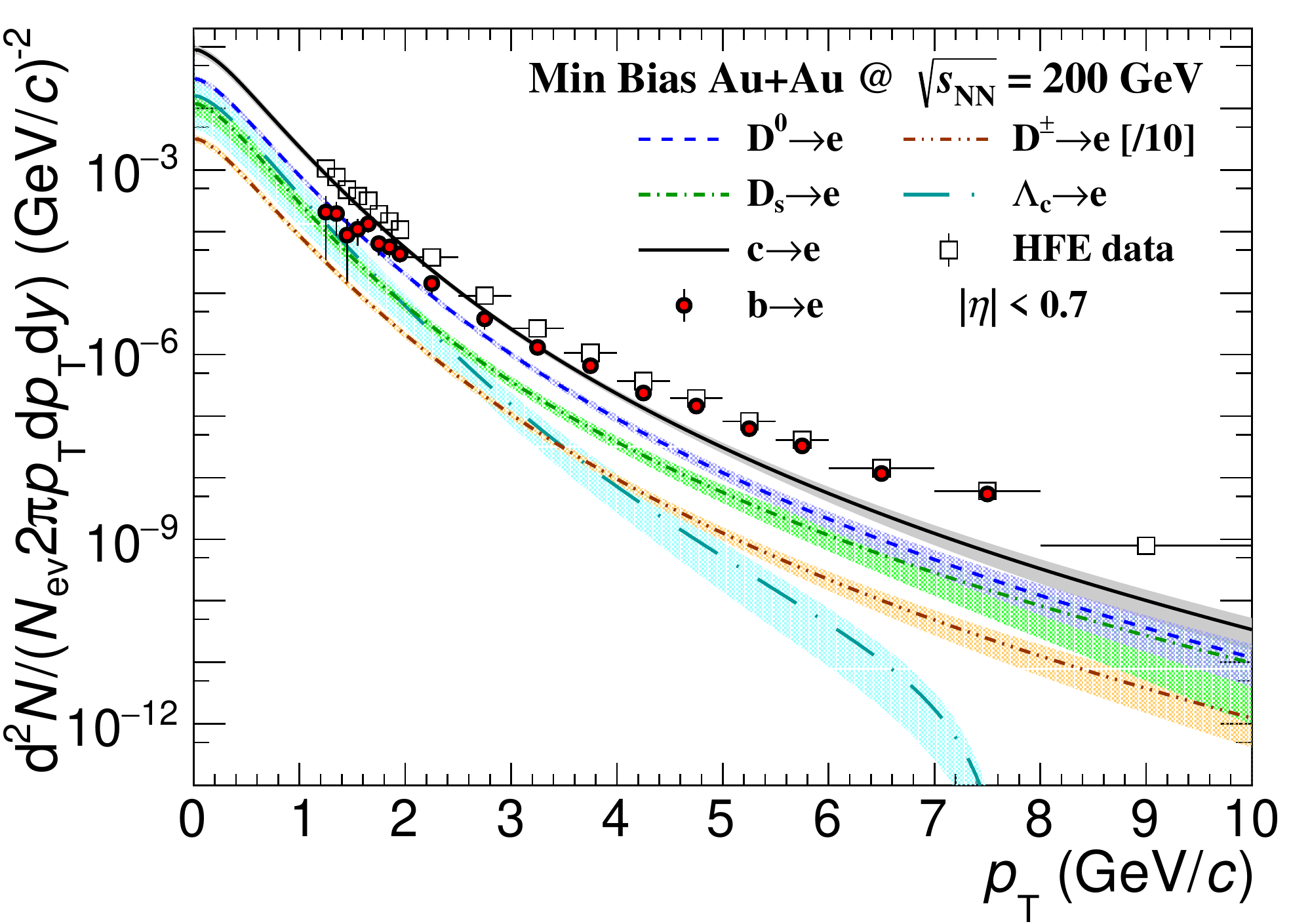}
\vspace{-2ex}
\caption{(Color online) 
Electron spectra from charmed hadrons (\dz~\cite{STARD1, STARD3}, \dpm~\cite{zhou2017measurements}, \ds~\cite{zhou2017measurements}, \lc~\cite{Lcspectra} and the sum of them ($\cte$)) and the inclusive HFE spectrum~\cite{HFE} at mid-rapidity ($\left|\eta\right|$ $<$ 0.7) in minimum bias Au+Au collisions at \sNN = 200 GeV. The spectrum of beauty hadron decay electrons ($\bte$) is obtained by subtracting the $\cte$ contributions from the HFE data. Uncertainties are shown as shaded bands.}
\label{fig1}
\end{figure}

Above open charm hadrons are simulated to decay to electrons via semileptonic decay channels with their parameterized \pT\ spectra and the Gaussian rapidity distribution ($\mu$ = 0 and $\sigma$ = 1.7) checked by the PYTHIA~\cite{pythia} event generator as inputs. The decay formfactors in the hadron rest frame are sampled from the measured distribution~\cite{Dformfactor}. As another check, the input charmed hadron rapidity distributions with scanning the standard deviation in a range of 1.4 $\leq$ $\sigma$ $\leq$ 2.0 result in little variation ($\lesssim$ 1\%) of the decay electron \pT\ spectra. Figure~\ref{fig1} shows the electron spectra from \dz\ (blue dashed curve), \dpm\ (brown dot-dot-dashed curve, scaled by 1/10), \ds\ (green dot-dashed curve) and \lc\ (cyan long-dot-dashed curve) decays and the summed charm contributions ($\cte$, black solid curve) at mid-rapidity ($\left|\eta\right|$ $<$ 0.7) in minimum bias Au+Au collisions at \sNN = 200 GeV. The electron spectra from charmed hadron decays are normalized by measured parent particle cross sections and semileptonic decay branching ratios~\cite{pdg}. The uncertainties of the charmed hadron \pT\ inputs are propagated into the decay electron spectra. The uncertainties of branching ratios are also taken into account. In particular, the uncertainty of $\rm D^{\pm}/D^0$ ratio is propagated into the $\rm D^{\pm}\rightarrow e$ spectrum. Electron spectra from \ds\ and \lc\ decays are scaled by $N_{\rm coll}$ to 0$-$80\% centrality from 10$-$40\% and 10$-$80\%, respectively, and the normalization uncertainties are counted. The total uncertainties of electron spectra from individual charmed hadron decays are shown as shaded bands in Fig.~\ref{fig1}. Uncertainty components of $\cte$ within 0 $-$ 10 \GeVc\ are summarized in Table~\ref{tab1}. As an example, with $\delta$ ($\sigma$) representing the relative (absolute) uncertainty, $\delta_{\rm D_s\rightarrow e}$ contributes $\left.\sigma_{\rm D_s\rightarrow e}\middle/\left(\cte\right)\right.=\delta_{\rm D_s\rightarrow e}\left[\left(\rm D_s\rightarrow e\right)\middle/\left(\cte\right)\right]$. Other uncertainty components are obtained in the same way. The total uncertainty of $\cte$ (11.0\%$-$54.7\%) is quadratically summed from all uncorrelated components. The black open squares denote the inclusive HFE spectrum ($\left|\eta\right|$ $<$ 0.7) measured by STAR~\cite{HFE}. The electron spectrum from beauty hadron decays ($\bte$), shown as red solid circles, is then calculated by subtracting the $\cte$ contribution from the inclusive HFE spectrum from \pT\ = 1.2 \GeVc\ to 8.0 \GeVc. In this and all of the following uncertainty calculations, statistical and systematic uncertainties are quadratically summed into total uncertainties, unless otherwise specified. The yield of $\rm D^{0}\rightarrow e$ at $p_{\rm T}^{\rm e}$ = 7.5 \GeVc\ from \dz\ decays at $p_{\rm T}^{\rm D^0}$ > 10 \GeVc\ contributes 52.8\% to the total electron yield in this $p_{\rm T}^{\rm e}$ bin. This fraction decreases to 26.2\% at $p_{\rm T}^{\rm e}$ = 6.5 \GeVc\ and becomes negligible at lower $p_{\rm T}^{\rm e}$. Based on this, the uncertainty of the last point of $\bte$ at \pT\ = 7.5 \GeVc\ is quoted conservatively with the 2-$\sigma$ uncertainty from $\cte$ due to higher \pT\ extrapolation.

\begin{table}
\centering
\caption{Uncertainty components of the $\cte$ spectrum (0 $-$ 10 \GeVc) from electron spectra from individual charmed hadron decays.}
\label{tab1}
\begin{threeparttable}
\resizebox{\linewidth}{!}{
\begin{tabular}{ccccc}
\toprule
                               & from input   & branching ratio & $N_{\rm coll}$ ratio & $\rm D^\pm/D^0$ \\
\midrule
$\delta\left(\rm D^{0}\rightarrow e\right)$       & 3.9\%$-$23.3\%\tnote{*} & 0.5\%$-$0.7\% & $\times$    & $\times$        \\
$\delta\left(\rm D^{\pm}\rightarrow e\right)$     & 4.0\%$-$24.2\%\tnote{*} & 0.6\%$-$0.8\% & $\times$    & 3.0\%$-$3.7\% \\
$\delta\left(\rm D_{s}\rightarrow e\right)$       & 2.1\%$-$27.1\% & 0.7\%$-$1.8\% & 1.2\%$-$3.0\% & $\times$    \\
$\delta\left(\rm \Lambda_{c}\rightarrow e\right)$ & $<$ 11.0\%     & $<$ 6.9\%     & $<$ 2.9\%     & $\times$    \\
\bottomrule
\end{tabular}}
\begin{tablenotes}
\footnotesize
\item[*] correlated
\end{tablenotes}
\end{threeparttable}
\end{table}

\begin{figure}
\centering
\includegraphics[width=\linewidth]{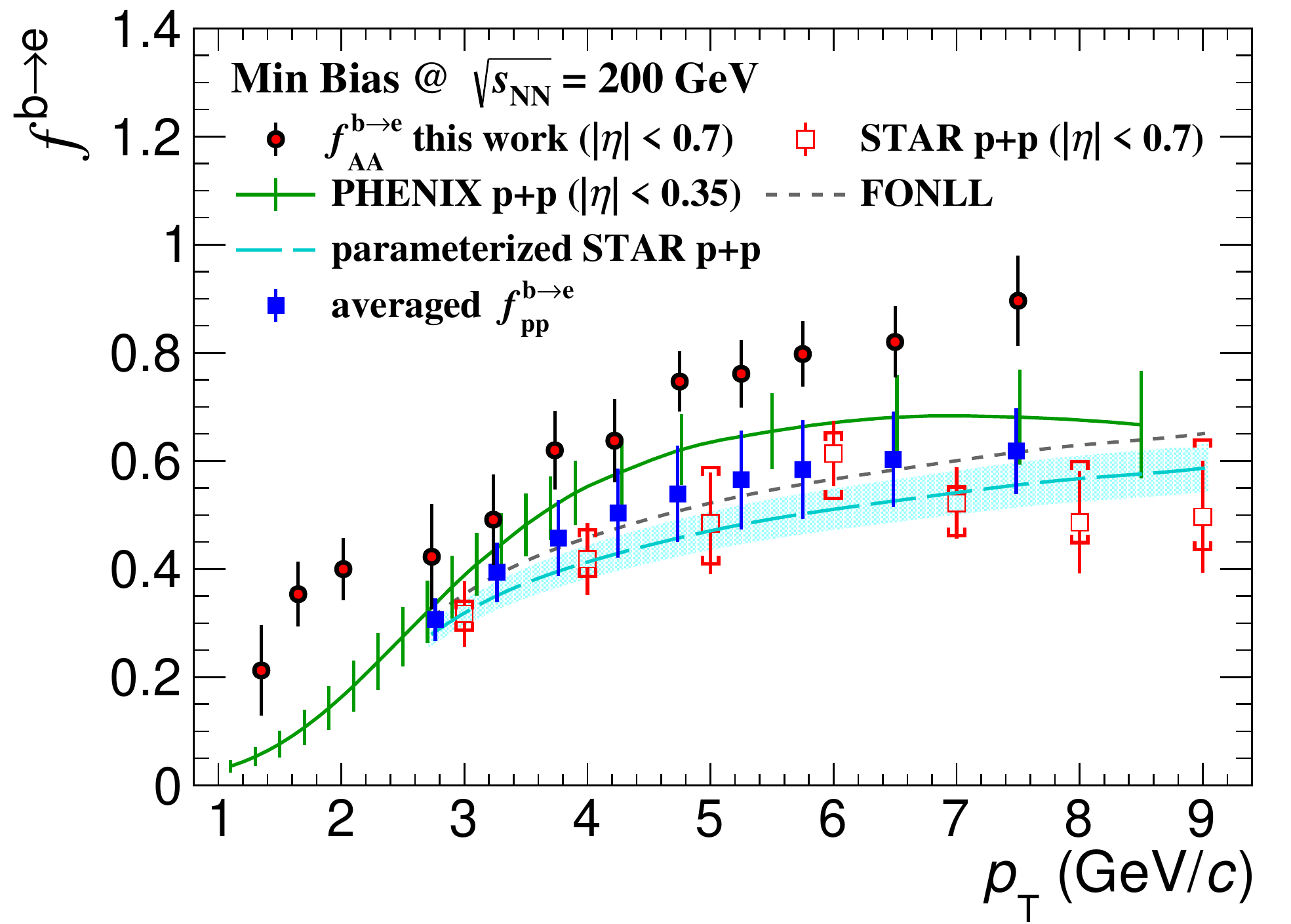}
\vspace{-2ex} 
\caption{(Color online) 
Beauty hadron decay electron fractions ($f^{\bte}$) at mid-rapidity ($\left|\eta\right|$ $<$ 0.7) in minimum bias Au+Au collisions compared with FONLL theoretical calculation~\cite{charm}, STAR ($\left|\eta\right|$ $<$ 0.7)~\cite{STARpp} and PHENIX ($\left|\eta\right|$ $<$ 0.35)~\cite{PHENIXpp} in $p$+$p$ collisions at \sNN = 200 GeV. $\fpp^{\bte}$ is the average of the parameterized STAR $p$+$p$ and the PHENIX $p$+$p$ data.}
\label{fig2}
\end{figure}

The beauty contribution fraction in the inclusive HFE spectrum in Au+Au collisions ($\fAA^{\bte}$) can be obtained by taking the ratio of the $\bte$ and HFE spectra, shown as solid circles in Fig.~\ref{fig2}. The relative uncertainty ($\delta$) of $\fAA^{\bte}$ (7.4\%$-$39.2\%) is propagated from those of the $\cte$ and HFE spectra with
\begin{equation}
\delta_{\fAA^{\bte}}^2=f_{\rm cb}^2\left(\delta_{\cte}^2+\delta_{\rm HFE}^2\right), 
\label{equ1}
\end{equation}
where $f_{\rm cb}=\left.\left(1-\fAA^{\bte}\right)\middle/\fAA^{\bte}\right.$, and the uncertainty of the last point at \pT\ = 7.5 \GeVc\ is quoted with the 2-$\sigma$ uncertainty from $\cte$ with the same approach as the uncertainty propagation of $\bte$. As a result, $\delta_{\cte}$ ($\delta_{\rm HFE}$) contributes 6.1\%$-$30.5\% (1.9\%$-$24.7\%). Here we compare the results with previous measurements in $p$+$p$ collisions by STAR ($\left|\eta\right|$ $<$ 0.7) via an electron-hadron correlation approach (red open squares)~\cite{STARpp} and by PHENIX ($\left|\eta\right|$ $<$ 0.35) with recent built-in vertex detector (green crosses)~\cite{PHENIXpp}. The FONLL calculation~\cite{charm} is presented as the gray dashed curve. The STAR $p$+$p$ data are fitted with the fixed FONLL function multiplied by a free parameter, which is shown as the cyan dashed curve with the band representing the uncertainty of the parameter given by the fit. The averaged $\fpp^{\bte}$ (blue solid squares) denotes the average of the parameterized STAR $p$+$p$ and the PHENIX $p$+$p$ data with half of their difference (contributes 8.1\%$-$15.1\%) and halves of their individual uncertainties (contribute 3.1\%$-$3.5\% and 6.0\%$-$9.5\%) quadratically summed into the uncertainty bars (12.8\%$-$16.5\%). The beauty contribution in the inclusive HFE spectrum in Au+Au collisions is clearly modified compared with that in $p$+$p$ collisions. At \pT\ $\sim$ 3.5 \GeVc, beauty and charm contributions are comparable, and at \pT\ $\sim$ 7.5 \GeVc\ the beauty contribution is up to 90\%, which is significantly higher than that in $p$+$p$ collisions. Since charm quark is strongly suppressed, the enhanced beauty fraction is consistent with less beauty quark suppression compared to charm quark in Au+Au collisions at \sNN = 200 GeV.

\begin{figure}
\centering
\includegraphics[width=\linewidth]{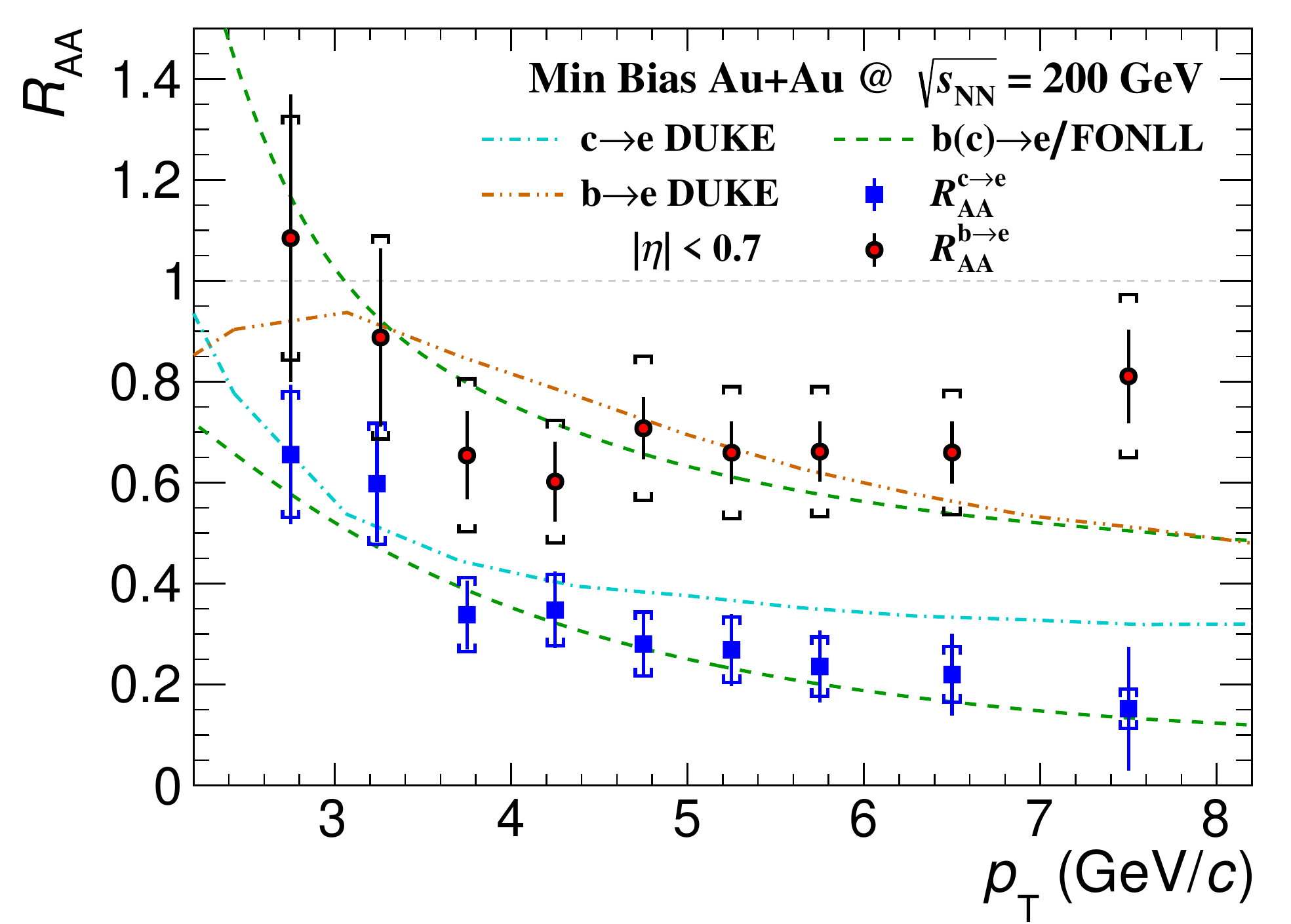}
\vspace{-2ex} 
\caption{(Color online) 
The nuclear modification factors ($\RAA$) of $\cte$ and $\bte$ at mid-rapidity ($\left|\eta\right|$ $<$ 0.7) in minimum bias Au+Au collisions at \sNN = 200 GeV. The solid bars represent the total uncertainties in Au+Au collisions and brackets denote those in $p$+$p$ collisions. Ratios of $\rm b(c)\rightarrow e/FONLL$ represented by dashed curves are obtained directly by the definition of $\RAA$. Results from DUKE model predictions~\cite{DUKE} are shown for comparison.}
\label{fig3}
\end{figure}

The $\RAA$ of electrons from individual charmed and beauty hadron decays ($\RAA^{\cte}$ and $\RAA^{\bte}$) can be extracted by
\begin{align}
\label{eq1}
\RAA^{\cte}=&\frac{1-\fAA^{\bte}}{1-\fpp^{\bte}}\RAA^{\rm ince},\\
\RAA^{\bte}=&\frac{\fAA^{\bte}}{\fpp^{\bte}}\RAA^{\rm ince},
\label{eq2}
\end{align}
where $\fAA^{\bte}$ is the beauty fraction in Au+Au collisions and $\fpp^{\bte}$ is the averaged beauty fraction of the parameterized STAR and the PHENIX data in $p$+$p$ collisions in Fig.~\ref{fig2}. The $\RAA^{\rm ince}$ is the $\RAA$ of inclusive electrons from HF hadron decays ($\left|\eta\right|$ $<$ 0.7) measured by STAR (Run 14)~\cite{HFERAA}. Figure~\ref{fig3} shows the $\RAA^{\cte}$ and $\RAA^{\bte}$ as functions of \pT\ extracted from Eq.~\eqref{eq1} and~\eqref{eq2} as blue squares and red circles, respectively. The $\RAA^{\bte}$ result is roughly consistent with the DUKE model prediction~\cite{DUKE}, and the latter predicts higher value than the $\RAA^{\cte}$ data at higher \pT. Two dashed curves representing $\rm b(c)\rightarrow e/FONLL$ are obtained directly by the definition of $\RAA$ as the parameterized spectra of $\rm b(c)\rightarrow e$ in Fig.~\ref{fig1} divided by their respective spectra from FONLL calculations~\cite{charm} scaled by $\langle N_{\rm coll}\rangle$ as a crosscheck, which shows a good agreement with data. Clear suppression at \pT\ $\gtrsim$ 3.5 \GeVc\ is observed for both $\RAA^{\cte}$ and $\RAA^{\bte}$, which indicates that charm and beauty quarks strongly interact with the hot-dense medium and lose energy. However, $\RAA^{\bte}$ shows less suppression compared with $\RAA^{\cte}$ at \pT\ $>$ 3.5 \GeVc, which is consistent with the mass-dependent energy loss prediction that beauty quark loses less energy due to the suppressed gluon radiation and smaller collisional energy exchange with the medium by its three-time larger mass compared to charm quark~\cite{Qeloss1, Qeloss2, Qeloss3}.

From Eq.~\eqref{eq1} and~\eqref{eq2}, the relative uncertainties ($\delta$) of $\RAA^{\cte}$ and $\RAA^{\bte}$ can be calculated by
\begin{align}
\label{equ2}
\delta_{\RAA^{\cte}}^2=&f_{\rm bc,AA}^2\delta_{\fAA^{\bte}}^2+f_{\rm bc,pp}^2\delta_{\fpp^{\bte}}^2+\delta_{\RAA^{\rm ince}}^2,\\
\delta_{\RAA^{\bte}}^2=&\delta_{\fAA^{\bte}}^2+\delta_{\fpp^{\bte}}^2+\delta_{\RAA^{\rm ince}}^2, 
\label{equ3}
\end{align}
respectively, where $f_{\rm bc}=\left.f^{\bte}\middle/\left(1-f^{\bte}\right)\right.$ in Au+Au or $p$+$p$ collisions. Table~\ref{tab2} summarizes the uncertainty components of $\RAA^{\cte}$ and $\RAA^{\bte}$.

\begin{table}
\centering
\caption{Uncertainty components of $\RAA^{\cte}$ and $\RAA^{\bte}$}
\label{tab2}
\resizebox{\linewidth}{!}{
\begin{tabular}{ccccc}
\toprule
              & $\fAA^{\bte}$ & $\fpp^{\bte}$ & $\RAA^{\rm ince}$ & total \\
\midrule
$\delta\left(\RAA^{\cte}\right)$ & 16.4\%$-$80.8\% & 5.7\%$-$22.3\%  & \multirow{2}*{12.4\%$-$22.1\%} & 27.8\%$-$85.1\% \\
$\delta\left(\RAA^{\bte}\right)$ & 7.4\%$-$23.1\%  & 12.8\%$-$16.5\% & ~                                & 20.8\%$-$34.5\% \\
\bottomrule
\end{tabular}}
\end{table}

\subsection{Elliptic Flow $v_2$}

The $v_2$ of \dz\ in 0$-$80\% Au+Au collisions at \sNN = 200 GeV measured by STAR~\cite{D0v2} is parameterized and extrapolated up to \pT\ = 10 \GeVc\ with a semi-empirical function as Eq.~\eqref{eq3}, which is modified from~\cite{v2func} with adding a linear term forced to pass through the origin according to the natural properties of $v_2$ and follow the number-of-constituent-quark (NCQ) scaling~\cite{v2func,NCQ} as $\left.v_2\middle/n\right.$ vs. $\left.\left(m_{\rm T}-m_0\right)\middle/n\right.$,
\begin{equation}
v_2=\frac{p_0n}{1+{\rm exp}\left(\frac{p_1-\left.\left(m_{\rm T}-m_0\right)\middle/n\right.}{p_2}\right)}-\frac{p_0n}{1+{\rm exp}\left(\frac{p_1}{p_2}\right)}+p_3\left(m_{\rm T}-m_0\right),
\label{eq3}
\end{equation}
where $m_{\rm T}=\sqrt{p_{\rm T}^2+m_0^2}$ and $m_0$ denote the transverse and rest masses of the particle, $n$ is the number of constituent quarks and $p_i$ ($i$ = 0, 1, 2, 3) are free parameters. At \pT\ above 10 GeV/$c$, the \dz\ $v_2$ is assumed as zero since due to the known effect of energy loss at high \pT, the particle $v_2$ drops quickly close to zero which has been observed in light particle~\cite{PHENIXpi0v2} and D meson~\cite{ALICEDv2} $v_2$ measurements in 200 GeV Au+Au and 2.76 TeV Pb+Pb collisions, respectively. Assuming \lc\ ($n$ = 3) also follows NCQ scaling as D-mesons ($n$ = 2), the $v_2$ of charmed mesons and baryons as functions of \pT\ can be obtained from Eq.~\eqref{eq3}. The azimuthal angle ($\phi$) distributions of charmed hadrons in each \pT\ bin follow the function~\cite{v2func} 
\begin{equation}
\frac{{\rm d}N}{{\rm d}\phi}=1+2v_2{\rm cos}(2\phi).
\label{eq4}
\end{equation}
Then the azimuthal distributions of electrons can be obtained by semileptonic decay simulations of charmed hadrons with their \pT\ spectra and azimuthal distributions as inputs. The $v_2$ of $\rm D\rightarrow e$ ($v_2^{\rm D\rightarrow e}$) and the $v_2$ of $\rm \Lambda_c\rightarrow e$ ($v_2^{\rm \Lambda_c\rightarrow e}$) can be obtained by fitting their azimuthal distributions with Eq.~\eqref{eq4} in each electron \pT\ bin.

The $v_2$ of $\cte$ ($v_2^{\cte}$) is an average of $v_2^{\rm D\rightarrow e}$ and $v_2^{\rm \Lambda_c\rightarrow e}$ with their relative yields as weights. In a similar way, the $v_2$ of $\bte$ ($v_2^{\bte}$) can be extracted from 
\begin{equation}
v_2^{\bte}=\frac{v_2^{\rm ince}-\left(1-\fAA^{\bte}\right)v_2^{\cte}}{\fAA^{\bte}}, 
\label{eq5}
\end{equation}
where $v_2^{\rm ince}$ denotes the $v_2$ of inclusive HFE from the parameterized average of the measurements by STAR~\cite{STARHFEv2} and PHENIX~\cite{PHENIXHFEv2} with Eq.~\eqref{eq3}. Note that the centralities for the STAR \dz\ $v_2$ input (0-80\%), the STAR HFE $v_2$ (0-60\%) and the PHENIX HFE $v_2$ (0-92\%) are different. Since the $v_2$ in minimum bias collisions is an average of $v_2$ values weighted by the particle yield in each centrality interval, one would expect the $v_2$ in minimum bias collisions should not be affected by the most peripheral collisions due to most of the particle yields coming from 0-60\% centrality. In fact, within current precision, HFE $v_2$ results measured from both STAR and PHENIX are consistent with each other.

The uncertainty from the parameterization of the \dz\ $v_2$ with the quadratic sum of statistical and systematic uncertainties, including the high \pT\ extrapolation, is propagated into the uncertainties of $v_2^{\rm D\rightarrow e}$ and $v_2^{\rm \Lambda_c\rightarrow e}$ through the decay simulations. The absolute uncertainty ($\sigma$) of $v_2^{\cte}$ with four components from $v_2$ and \pT\ spectra of $\rm D\rightarrow e$ and $\rm \Lambda_c\rightarrow e$ can be obtained with the variant of the differentiated $v_2^{\cte}$ calculation formula as
\begin{equation}
\begin{split}
\sigma_{v_2^{\cte}}^2=&\left(f_{\rm Dc}\sigma_{v_2^{\rm D\rightarrow e}}+f_{\rm \Lambda_cc}\sigma_{v_2^{\rm \Lambda_c\rightarrow e}}\right)^2\\
&+\left(\frac{\Delta v_2}{\cte}\right)^2\left(f_{\rm \Lambda_cc}^2\sigma_{\rm D\rightarrow e}^2+f_{\rm Dc}^2\sigma_{\rm \Lambda_c\rightarrow e}^2\right),
\end{split}
\label{equ4}
\end{equation}
where $f_{\rm Dc}=\left.\left(\rm D\rightarrow e\right)\middle/\left(\cte\right)\right.$, $f_{\rm \Lambda_cc}=\left.\left(\rm \Lambda_c\rightarrow e\right)\middle/\left(\cte\right)\right.$ and $\Delta v_2=v_2^{\rm D\rightarrow e}-v_2^{\rm \Lambda_c\rightarrow e}$. Uncertainties of $v_2^{\rm D\rightarrow e}$ and $v_2^{\rm \Lambda_c\rightarrow e}$ are correlated, since both of them are propagated from the uncertainty of \dz\ $v_2$. From Eq.~\eqref{eq5}, we can calculate three parts of the absolute uncertainty ($\sigma$) of $v_2^{\bte}$ with
\begin{equation}
\sigma_{v_2^{\bte}}^2=f_{\rm 1b}^2\sigma_{v_2^{\rm ince}}^2+f_{\rm cb}^2\sigma_{v_2^{\cte}}^2+f_{\rm 1b}^4\left(v_2^{\cte}-v_2^{\rm ince}\right)^2\sigma_{\fAA^{\bte}}^2,
\label{equ5}
\end{equation}
where $f_{\rm 1b}=\left.1\middle/\fAA^{\bte}\right.$ and $f_{\rm cb}=\left.\left(1-\fAA^{\bte}\right)\middle/\fAA^{\bte}\right.$.
Each of the three parts is strongly controlled by $\fAA^{\bte}$ and the low values of $\fAA^{\bte}$ at low \pT\ result in large uncertainties of $v_2^{\bte}$.

\begin{figure}
\centering
\includegraphics[width=\linewidth]{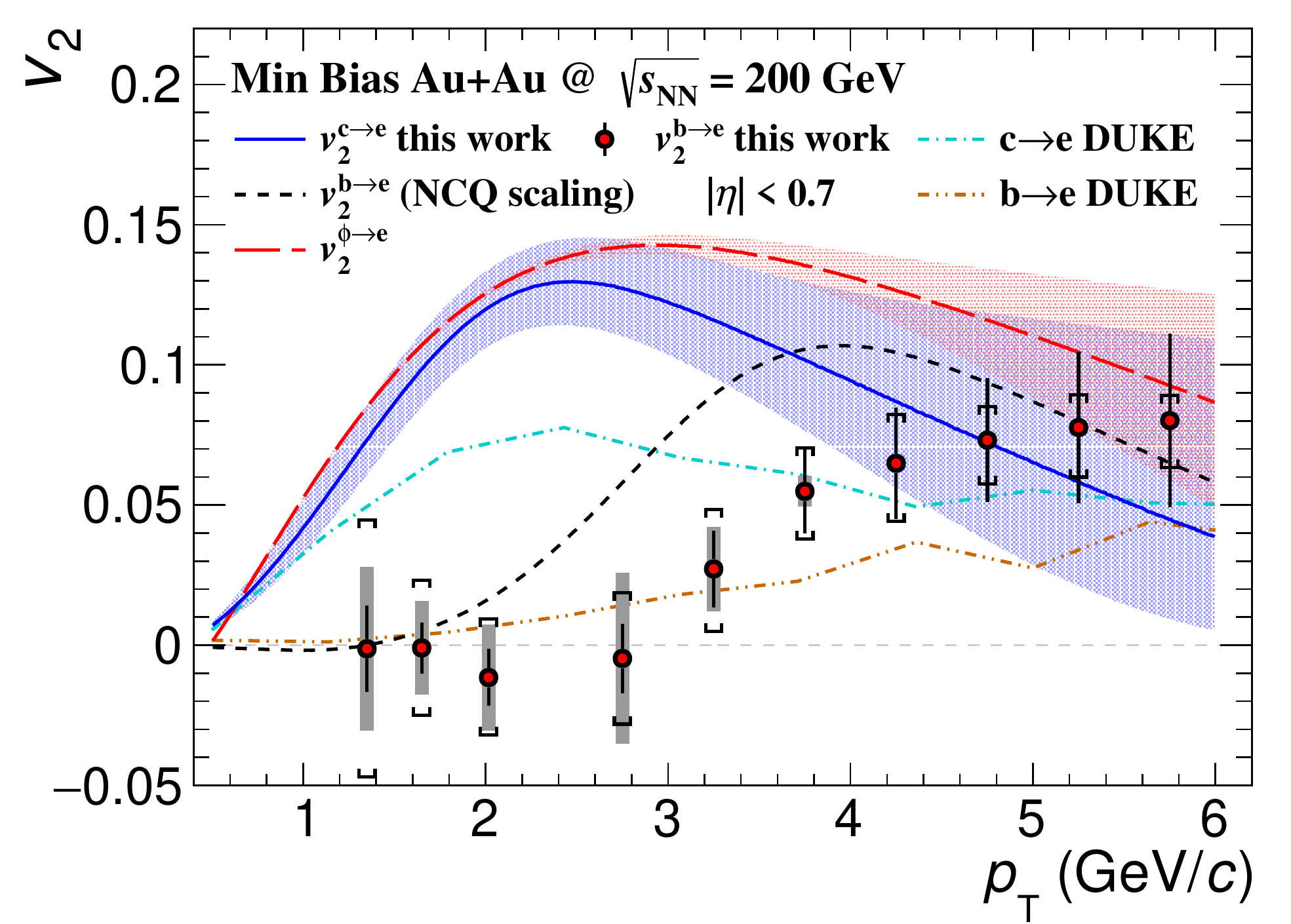}
\vspace{-2ex} 
\caption{(Color online) 
The elliptic flows ($v_2$) of $\cte$ and $\bte$ at mid-rapidity ($\left|\eta\right|$ $<$ 0.7) in minimum bias Au+Au collisions at \sNN = 200 GeV. The major contributions to the uncertainty of $v_2^{\bte}$ are from $v_2^{\rm ince}$ (bars), $v_2^{\cte}$ (brackets) and $\fAA^{\bte}$ (grey bands). The $v_2^{\rm b\rightarrow e}$ with B-meson $v_2$ NCQ scaling assumption and the $v_2^{\rm \phi\rightarrow e}$ are shown as the dashed curve and open squares, respectively. Results from DUKE~\cite{DUKE} model predictions are shown for comparison.}
\label{fig4}
\end{figure}

Figure~\ref{fig4} shows the results of $v_2^{\cte}$ and $v_2^{\bte}$ as the blue solid curve with an uncertainty band and red circles, respectively. The $v_2$ of $\rm\phi\rightarrow e$ ($v_2^{\rm\phi\rightarrow e}$, red long-dashed curve with band) is obtained in the same way as $v_2^{\rm D\rightarrow e}$ with the $\rm\phi$-meson spectrum~\cite{phispectrum} and $v_2$ (0$-$80\%)~\cite{phiv2} as inputs. DUKE model predictions~\cite{DUKE} are also shown as dot-dashed curves for comparison. The electron $v_2$ from beauty hadron decays at \pT\ $>$ 3.0 \GeVc\ is observed with an average of 4-sigma significance ($\chi^2/ndf$ = 29.7/6) deviating from zero. And it is consistent with electrons from charmed or strange hadron decays within uncertainties at \pT\ $>$ 4.5 \GeVc. This flavor independent $v_2$ at high \pT\ could be attributed to the initial geometry anisotropy or the path length dependence of the energy loss in the medium. A smaller $v_2^{\bte}$ compared with $v_2^{\cte}$ is observed at \pT\ $<$ 4.0 \GeVc, which may be driven by the larger mass of beauty quark than that of charm quark. The black dashed curve represents the $v_2^{\rm b\rightarrow e}$ assuming that B-meson $v_2$ follows the NCQ scaling, which is from the same technique as $v_2^{\rm D\rightarrow e}$ with a decay formfactor in the B-meson frame sampled from the distribution measured by CLEO~\cite{Bformfactor}. The $v_2^{\bte}$ presented here, as a mixture of $v_2$ of beauty and light quarks via a coalescence hadronization, deviates from the curve at 2.5 \GeVc\ $<$ \pT\ $<$ 4.5 \GeVc\ with a confidence level of 99\% ($\chi^2/ndf$ = 14.3/4), which favors that the beauty quark elliptic flow is smaller than that of light quarks, unlike the \dz\ $v_2$ scaled with that of light flavor hadrons by dividing number of constituent quarks in both $v_2$ and $\left(m_{\rm T}-m_0\right)$~\cite{D0v2}. This suggests that beauty quark is unlikely thermalized and too heavy to be moved following the collective flow of lighter partons. 

\section{Summary}
In summary, this paper reports the individual electron \pT\ spectra, $\RAA$ and $v_2$ distributions from charmed and beauty hadron decays in minimum bias Au+Au collisions at \sNN = 200 GeV at RHIC. We find that the electron $\RAA$ from beauty hadron decays is suppressed at high \pT\ $>$ 3.5 \GeVc\, but less suppressed compared with that from charmed hadron decays, which indicates that beauty quark interacts with the hot-dense medium and loses energy and is consistent with the mass-dependent energy loss scenario. For the first time, the non-zero electron $v_2$ from beauty hadron decays at \pT\ $>$ 3.0 \GeVc\ is observed and consistent with hadrons containing charm or strangeness at \pT\ $>$ 4.5 \GeVc\ with large uncertainties, which could be mainly due to the initial geometry anisotropy or the path length dependence of the energy loss in the medium. And its smaller elliptic flow compared with that from charmed hadron decays at \pT\ $<$ 4.0 \GeVc\ is observed. At 2.5 \GeVc\ $<$ \pT\ $<$ 4.5 \GeVc, $v_2^{\bte}$ is smaller than a number-of-constituent-quark scaling hypothesis, which suggests that the extremely heavy mass of beauty quark prevents itself participating in the partonic collectivity and the first non-thermalized particle (beauty quark) is observed in heavy-ion collisions at RHIC energy.

\textit{Acknowledgments:} The authors would like to thank Guannan Xie, Zebo Tang, Rongrong Ma, Guangyou Qin and Hans Georg Ritter for constructive discussions. This work was supported by National Key R\&D Program of China with Grant No.~2018YFE0205200, Strategic Priority Research Program of Chinese Academy of Sciences with Grant No.~XDB34000000, National Natural Science Foundation of China with Grant No.~11890712 and Anhui Provincial Natural Science Foundation with Grant No.~1808085J02.


\begin{thebibliography}{99}
\bibitem{QGP1} M. Gyulassy, arXiv:nucl-th/0403032 (2004).

\bibitem{QGP2} S.A. Bass, M. Gyulassy, H. Stoecker and W. Greiner, J. Phys. G \textbf{25}, R1-R57 (1999).

\bibitem{QGP3} F. Karsch, Nucl. Phys. A \textbf{590}, 367c-382c (1995).

\bibitem{QGPexp1} J. Adams {\it et al.} (STAR Collaboration), Nucl. Phys. A \textbf{757}, 102-183 (2005).

\bibitem{QGPexp2} K. Adcox {\it et al.} (PHENIX Collaboration), Nucl. Phys. A \textbf{757}, 184-283 (2005).

\bibitem{QGPexp3} B. Muller, J. Schukraft and B. Wyslouch, Ann. Rev. Nucl. Part. Sci. \textbf{62}, 361-386 (2012).

\bibitem{QGPexp4} Y. Akiba {\it et al.}, arXiv:1502.02730 (2015).

\bibitem{BM} B. Muller, Nucl. Phys. A \textbf{750}, 84-97 (2005).

\bibitem{charm} M. Cacciari, P. Nason and R. Vogt, Phys. Rev. Lett. \textbf{95}, 122001 (2005).

\bibitem{Qeloss1} Y.L. Dokshitzer and D.E. Kharzeev, Phys. Lett. B \textbf{519}, 199-206 (2001).

\bibitem{Qeloss2} M. Djordjevic, M. Gyulassy and S. Wicks, Phys. Rev. Lett. \textbf{94}, 112301 (2005).

\bibitem{Qeloss3} A. Buzzatti and M. Gyulassy, Phys. Rev. Lett. \textbf{108}, 022301 (2012).

\bibitem{RAAXNW} X.-N. Wang and M. Gyulassy, Phys. Rev. Lett. \textbf{68}, 1480 (1992).

\bibitem{STARD1} L. Adamczyk {\it et al.} (STAR Collaboration), Phys. Rev. Lett. \textbf{113}, 142301 (2014); Phys. Rev. Lett. \textbf{121}, 229901 (2018).

\bibitem{STARD3} J. Adam {\it et al.} (STAR Collaboration), Phys. Rev. C \textbf{99}, 034908 (2019).

\bibitem{ExpE1} B.I. Abelev {\it et al.} (STAR Collaboration), Phys. Rev. Lett. \textbf{98}, 192301 (2007); Phys. Rev. Lett. 106 159902 (2011).

\bibitem{ExpE2} A. Adare {\it et al.} (PHENIX Collaboration), Phys. Rev. Lett. \textbf{98}, 172301 (2007).

\bibitem{PhecbE} A. Adare {\it et al.} (PHENIX Collaboration), Phys. Rev. C \textbf{93}, 034904 (2016).

\bibitem{HFTTemplate} K. Oh (STAR Collaboration), Nucl. Phys. A \textbf{967}, 632-635 (2017).

\bibitem{LHCb2e} J. Adam {\it et al.} (ALICE Collaboration), JHEP \textbf{07}, 052 (2017).

\bibitem{FlowArt} A.M. Poskanzer and S.A. Voloshin, Phys. Rev. C \textbf{58}, 1671 (1998).

\bibitem{Moore} G.D. Moore and D. Teaney, Phys. Rev. C \textbf{71}, 064904 (2005).

\bibitem{Andronic} A. Andronic {\it et al.}, Eur. Phys. J. C \textbf{76}, 107 (2016).

\bibitem{SUBATECH} M. Nahrgang, J. Aichelin, S. Bass, P.B. Gossiaux and K. Werner, Phys. Rev. C \textbf{91}, 014904 (2015).

\bibitem{TAMU1} M. He, R.J. Fries and R. Rapp, Phys. Rev. C \textbf{86}, 014903 (2012).

\bibitem{TAMU2} M. He, R.J. Fries and R. Rapp, Phys. Rev. Lett. \textbf{110}, 112301 (2013).

\bibitem{LBT} S. Cao, T. Luo, G.-Y. Qin and X.-N. Wang, Phys. Rev. C \textbf{94}, 014909 (2016).

\bibitem{PHSD} T. Song, H. Berrehrah, D. Cebrera, J.M. Torres-Rincon, L. Tolos, W. Cassing and E. Bratkovskaya, Phys. Rev. C \textbf{92}, 014910 (2015).

\bibitem{lQCD1} H. Ding, A. Francis, O. Kaczmarek, F. Karsch, H. Satz and W. Soeldner, Phys. Rev. D \textbf{86}, 014509 (2012).

\bibitem{lQCD2} D. Banerjee, S. Datta, R. Gavai and P. Majumdar, Phys. Rev. D \textbf{85}, 014510 (2012).

\bibitem{zhou2017measurements} L. Zhou (STAR Collaboration), Nucl. Phys. A \textbf{967}, 620-623 (2017).

\bibitem{Lcspectra} J. Adam {\it et al.} (STAR Collaboration), Phys. Rev. Lett. \textbf{118}, 212301 (2020); arXiv:1910.14628.

\bibitem{b2D} X. Chen (STAR Collaboration), PoS \textbf{345}, 158 (2019).

\bibitem{wilk2000interpretation} G. Wilk and Z. W{\l}odarczyk, Phys. Rev. Lett. \textbf{84}, 2770 (2000).

\bibitem{adamczyk2012measurements} L. Adamczyk {\it et al.} (STAR Collaboration), Phys. Rev. D \textbf{86}, 072013 (2012).

\bibitem{ghosh2014diffusion} S. Ghosh, S.K. Das, V. Greco, S. Sarkar and J.-e Alam, Phys. Rev. D \textbf{90}, 054018 (2014).

\bibitem{zhao2018sequential} J. Zhao, S. Shi, N. Xu and P. Zhuang, arXiv:1805.10858 (2018).

\bibitem{ko} S. Cho, K.-J. Sun, C.M. Ko, S.H. Lee and Y. Oh, arXiv:1905.09774 (2019).

\bibitem{rapp} M. He and R. Rapp, arXiv:1905.09216 (2019).

\bibitem{pythia} T. Sj\"ostrand, S. Mrenna and P. Skands, JHEP \textbf{05}, 026 (2006).

\bibitem{Dformfactor} H. Liu, Y. Zhang, C. Zhong and Z. Xu, Phys. Lett. B \textbf{639}, 441-446 (2006).

\bibitem{pdg} M. Tanabashi {\it et al.} (Particle Data Group), Phys. Rev. D \textbf{98}, 030001 (2018).

\bibitem{HFE} M. Mustafa (STAR Collaboration), Nucl. Phys. A \textbf{904-905}, 665c-668c (2013).

\bibitem{STARpp} M.M. Aggarwal {\it et al.} (STAR Collaboration), Phys. Rev. Lett. \textbf{105}, 202301 (2010).

\bibitem{PHENIXpp} C. Aidala {\it et al.} (PHENIX Collaboration), Phys. Rev. D \textbf{99}, 092003 (2019).

\bibitem{HFERAA} B.I. Abelev {\it et al.} (STAR Collaboration), Phys.Rev.Lett. 98, 192301 (2007) and Phys.Rev.Lett. 106, 159902 (2011) (erratum); Y. Zhang, G. Xie, S. Zhang and X. Dong, Sci. Sin.-Phys. Mech. Astron. \textbf{49}, 102003 (2019); S. Zhang. Measurements of electrons from open heavy flavor hadron decays in p+p and Au+Au collisions at $\sqrt{s_{\rm NN}}$ = 200 GeV by the STAR experiment. Ph.D thesis (2018), https://drupal.star.bnl.gov/STAR/theses/phd/shenghuizhang-0

\bibitem{DUKE} S. Cao, G.-Y. Qin and S.A. Bass, Phys. Rev. C \textbf{92}, 024907 (2015).

\bibitem{D0v2} L. Adamczyk {\it et al.} (STAR Collaboration), Phys. Rev. Lett. \textbf{118}, 212301 (2017).

\bibitem{v2func} X. Dong, S. Esumi, P. Sorensen, N. Xu and Z. Xu, Phys. Lett. B \textbf{597}, 328-332 (2004).

\bibitem{NCQ} J. Tian , J.-H. Chen, Y.-G. Ma, X.-Z. Cai, F. Jin, G.-L. Ma, S. Zhang and C. Zhong, Phys. Rev. C \textbf{79}, 067901 (2009).

\bibitem{PHENIXpi0v2} S.S. Adler {\it et al.} (PHENIX Collaboration), Phys. Rev. Lett. \textbf{96}, 032302 (2006).

\bibitem{ALICEDv2} B. Abelev {\it et al.} (ALICE Collaboration), Phys. Rev. Lett. \textbf{111}, 102301 (2013).

\bibitem{STARHFEv2} L. Adamczyk {\it et al.} (STAR Collaboration), Phys. Rev. C \textbf{95}, 034907 (2017).

\bibitem{PHENIXHFEv2} T. Hachiya (PHENIX Collaboration), Nucl. Phys. A \textbf{982}, 663-666 (2019).

\bibitem{phispectrum} B.I. Abelev {\it et al.} (STAR Collaboration), Phys. Rev. Lett. \textbf{99}, 112301 (2007).

\bibitem{phiv2} L. Adamczyk {\it et al.} (STAR Collaboration), Phys. Rev. Lett. \textbf{116}, 062301 (2016).

\bibitem{Bformfactor} V. Jain (CLEO Collaboration), Nucl. Phys. B (Proc. Suppl.) \textbf{50}, 96-102, (1996).


\end{thebibliography}
\end{document}